\documentstyle[12pt,graphicx]{article}

\begin{document}

\title{\bf The rational parts of one-loop QCD amplitudes II: The five-gluon case}

\author{Xun Su\thanks{E-mail: ghkcn@itp.ac.cn} $^{1}$,
Zhi-Guang Xiao\thanks{E-mail: zhgxiao@itp.ac.cn} $^{1,2}$, Gang
Yang\thanks{E-mail: yangg@itp.ac.cn} $^{1}$and Chuan-Jie
Zhu\thanks{E-mail: zhucj@itp.ac.cn} $^{1,3}$ }

\maketitle

\medskip
\centerline{$^1$Institute of Theoretical Physics, Chinese Academy
of Sciences} \centerline{P. O. Box 2735, Beijing 100080, P. R.
China}
\medskip
\centerline{$^2$The Interdisciplinary Center of Theoretical
Studies, Chinese} \centerline{Academy of Sciences, P. O. Box 2735,
Beijing 100080, P.~R.~China}
\medskip
\centerline{$^3$Center of Mathematical Science, Zhejiang
University} \centerline{Hangzhou 310027, P. R. China}

\begin{abstract}

The rational parts of 5-gluon one-loop amplitudes are computed by
using the newly developed method for computing the rational parts
directly from Feynman integrals. We found complete agreement with
the previously well-known results of Bern, Dixon and Kosower
obtained by using the string theory method. Intermediate results
for some combinations of Feynman diagrams are presented in order
to show the efficiency of the method and the local cancellation
between different contributions.

\end{abstract}
\newpage

\section{Introduction}
In a  previous paper \cite{xyzi}, we developed a method for
computing the rational parts of one-loop amplitudes directly
from Feynman integrals. The purpose of this paper is to apply
this method to compute the rational parts of 5-gluon one-loop
amplitudes. The result agrees with the well-known result of Bern,
Dixon and Kosower \cite{BDK} obtained first by using
string-inspired methods. (Other 5-parton amplitudes in massless QCD
were later computed by using either standard Feynman diagrammatic technique
\cite{KST} or supersymmetric decomposition and perturbative
unitarity \cite{BDKM}.)

The computation of the multi-particle one-loop amplitudes in QCD is
a very difficult problem. Even for 4-parton amplitudes the
computation is quite non-trivial \cite{EllisSexton}. For 5-gluon
amplitudes a new method was developed \cite{BernKosower87} by using
string theory.

The constant effort to calculate multi-leg one-loop amplitudes
lies in the application to the forthcoming experimental program at
CERN's Large Hadron Collider (LHC), as there are lots of processes
with many particles as final states \cite{Salam}. We refer the
reader to \cite{xyzi} for a discussion and extensive references
for the recent efforts in computing the multi-leg one-loop
amplitudes and the recent developments inspired by twistor string
theory \cite{Witten,CSW}.

In order to compute multi-leg one-loop amplitudes in QCD, it is a
good strategy \cite{BDDK} to decompose the QCD amplitudes into
simpler ones by using the supersymmetric decomposition:
\begin{equation}
A^{QCD} = A^{N=4} - 4 A^{N=1~{\rm chiral}} + A^{N=0~{\rm or~
scalar}} ,
\end{equation}
where $A^{QCD}$ denotes an amplitude with only a gluon circulating
in the loop, $A^{N=4,1}$ have the full $N=4,1$ multiplets
circulating in the loop, and $A^{N=0}$ has only a complex scalar
in the loop.

By using the general properties of the one-loop amplitudes, Bern,
Dunbar, Dixon and Kosower  proved that the supersymmetric
amplitudes  $A^{N=4,1}$ are completely determined by 4-dimensional
unitarity \cite{BDDK}, i.e. the amplitudes are completely
cut-constructible and the rational part is vanishing (see
\cite{BDDK,xyzi} for more detail explanation). For MHV helicity
configurations, explicit results were obtained for $A^{N=4}$ in
\cite{BDDK}. The recent development of using MHV vertices to
compute one-loop amplitudes leads to many new results for the
cut-constructible part
\cite{BST,Cachazo,RecentTwistorBern,Rozali,BBST,Dunbar,BernA,Luo,BCFW,
BoFengA,BoFengC,BoFengSix}. In particular, Bedford, Brandhuber,
Spence and  Travaglini \cite{BST,BBST} applied the MHV
vertices to one-loop calculations. Britto, Buchbinder, Cachazo,
Feng and Mastrolia \cite{BoFengA, BoFengC, BoFengSix} developed an
efficient technique for evaluating the rational coefficients in an
expansion of the one-loop amplitudes in terms of scalar box,
triangle and bubble integrals (the cut-constructible part, see
\cite{xyzi} for details). By using their technique, it is much
easier to calculate the coefficients of box integrals without doing
any integration. Recently, Britto, Feng and Mastrolia completed
the computation of the cut-constructible terms for all the 6-gluon
helicity amplitudes \cite{BoFengSix}.

In order to complete the QCD calculation for the 6-gluon helicity
amplitudes, the remaining challenge is to compute the rational
parts of the helicity amplitudes with scalars circulating in the
loop. In general, we need an efficient and powerful method to
compute the rational part of any amplitude.

As we reviewed in \cite{xyzi}, there are various approaches
\cite{BDKBX,BernMorgan,BDKR,AMST,BDKA, BDKB} to compute the
rational part. In particular, Bern, Dixon and Kosower \cite{BDKA,
BDKB} developed the bootstrap recursive approach which has lead
to quite general results \cite{FordeKosower,BDKC,BDKE}. In this
paper we will use the approach as developed in \cite{xyzi} and
apply it to compute the rational parts of the 5-gluon one-loop
amplitudes. In another paper \cite{xyziii} we will compute the
rational parts of 6-gluon amplitudes in QCD (see also
\cite{BDKE}) which are the last missing pieces for the complete
partial helicity amplitudes of the 6-gluon one-loop QCD amplitude.

This paper is organized as follows: in Sect.~2, we recall briefly
the Feynman diagrams and the Feynman rules, tailored for our
computation of the 5-gluon amplitudes. Some simple reduction
formulas are recalled briefly in Sect.~3. In Sect.~4 we summarize
all the integral formulas we will use in this paper. Then the
following 2 sections present the results for the rational parts of
the two independent MHV helicity configurations.

\section{Notation, the Feynman diagrams and the Feynman rules}

A word about notation: we use the same notation as given in
\cite{xyzi}. We use $\epsilon_{i(i+1)\cdots (i+n)}$ to denote the
composite polarization vectors for sewing trees to the loop.

For the purpose of this paper we consider only the Feynman diagrams
and Feynman rules for the one-loop gluon amplitudes with scalars
circulating in the loop. We do not follow the usual convention of
differentiating different particles by different kinds of lines
because there are only two kinds of particles: gluons and scalars,
and scalars only appear in the loop.

\begin{figure}[ht]
\centerline{\includegraphics[height=6cm]{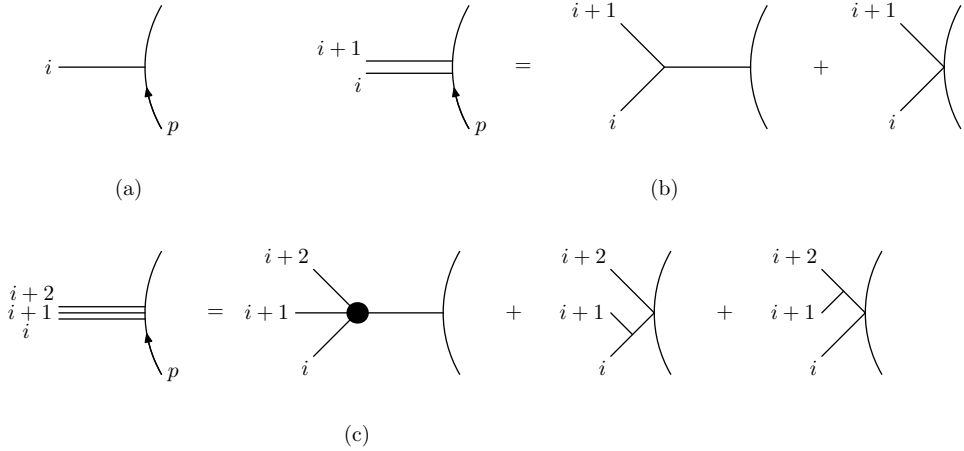} } \caption{The
Feynman rules for sewing trees to loop. The blob denotes an
expansion of the tree amplitude.} \label{FeynmanTree}
\end{figure}

For the explicit calculation of one-loop amplitudes by the usual
Feynman diagram technique, we can first collect all terms with the
same loop structure into one entity. Generally a few cyclicly
consecutive external lines are joined in tree diagrams and
connected to  the same point on the loop (sewing trees to the loop).
We denote the sum of all these contributions by $P_{i(i+1)\cdots
(i+m-1)}$ for $m$ such external lines. For $m=1,2,3$, the relevant
Feynman diagrams are shown in Fig.~\ref{FeynmanTree}. Explicitly
we have:
\begin{eqnarray}
P_{i}(p) & = & (\epsilon_i,p) = (\epsilon_{i},p-k_i), \\
P_{i(i+1)}(p) & = & (\epsilon_{i(i+1)},p) - {1\over2} \,
 (\epsilon_i,\epsilon_{i+1})  , \\
P_{i(i+1)(i+2)}(p)
 & = & ( \epsilon_{i(i+1)(i+2)},p) - {1\over2}
\, (
    (\epsilon_{i(i+1)},\epsilon_{i+2}) +  (\epsilon_i,\epsilon_{(
    i+1)(i+2)}) ) ,
\end{eqnarray}
where $\epsilon_{\cdots}$'s are composite polarization vectors
introduced in \cite{xyzi}. The computation of these composite
polarization vectors is a simplified version of the general
recursive calculation of the tree-level $n$-gluon amplitudes
\cite{BerendsGiele}.

\begin{figure}[ht]
\centerline{\includegraphics[height=5.5cm]{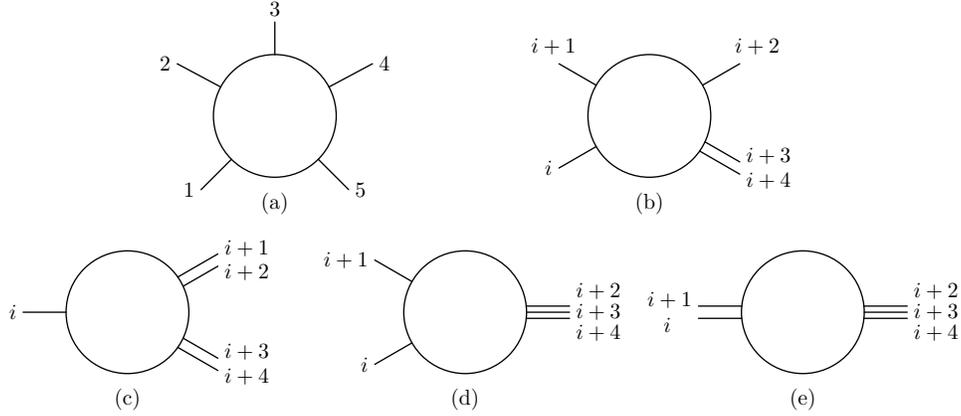} }
\caption{All the possible one-loop Feynman diagrams for 5 gluons.
The index $i$ runs from 1 to 5 if there is an index $i$. }
\label{FeynmanFivePoint}
\end{figure}

Considering all these different tree diagrams just as the same
diagram and denoting them by multiple parallel lines attached to
the loop, we have only 21 different Feynman diagrams for the
5-gluon one-loop amplitude (with scalars circulating the loop).
Some representative diagrams are given in
Fig.~\ref{FeynmanFivePoint}. The counting goes as follows:
\begin{itemize}
\item 1 pentagon diagram, the diagram (a);
\item 5 box diagrams because there are 5 different ways of combining
two consecutive external lines, diagram (b) with $i=1,\cdots,5$;
\item 10 triangle diagrams which are further divided into 5 two-mass
 triangle diagrams (diagram (c)) and 5 one-mass triangle
diagrams (diagram (d));
\item 5 bubble diagrams (e).
\end{itemize}

It is straightforward to compute the rational part from each diagram
for a given helicity configuration. In the next two sections we will
review all the formulas needed.

\section{Tensor reduction of the one-loop amplitudes}

There is a vast literature on this subject
\cite{PassarinoVeltman,Melrose:1965kb,BDKReduction,
Binothz,Tarasov,Duplancic:2003tv,Denner}.  The tensor reduction relations we will
use for our calculation of 5-gluon amplitudes are quite simple.
It is based on the BDK trick \cite{BDKB} of multiplying and
dividing by spinor square roots. For 5-gluon amplitudes it is not
so important to make a clever choice of reference momenta. For
6-gluon amplitudes it is important to make a specific choice of the
reference momenta to make all tensor reductions simple enough to
obtain relatively compact analytic results.

\begin{figure}[ht]
\centerline{\includegraphics[height=4cm]{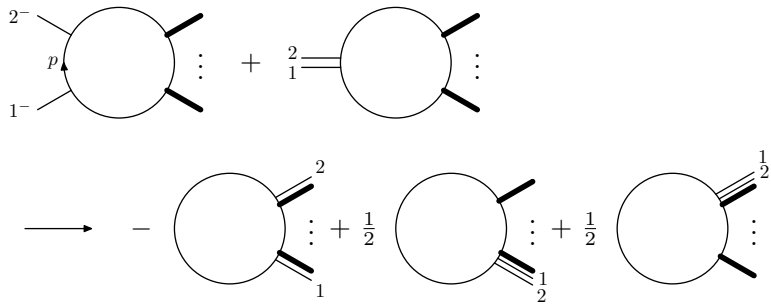} }
\caption{For two adjacent same helicities, the tensor reduction for
the combination of two diagrams is even simpler.} \label{Four}
\end{figure}

For the tensor reduction with only 2 neighboring same helicity
external gluons, it is possible to choose the reference momenta to
be each other's momenta, i.e.  $\epsilon_1 =
\lambda_1\tilde\lambda_2$, $\epsilon_2 =
\lambda_2\tilde\lambda_1$.\footnote{An overall constant is omitted
for the polarization vector which can be easily reinstated at the
end of calculation. See \cite{ChineseMagic,DixonReview} for
details.} The tensor reduction is done by considering the
contributions from 2 diagrams together and the result formula is
shown pictorially in Fig.~\ref{Four}. The exact algebraic formula
is:
\begin{eqnarray}
 {(\epsilon_1,p+k_1)(\epsilon_2,p ) \over (p+k_1)^2
 p^2(p-k_{2 })^2} &  + & { (\epsilon_{12}, p+k_1) -
(\epsilon_1,\epsilon_2)/2 \over (p+k_1)^2 (p-k_{2})^2}
\nonumber \\
& = & - {1\over p^2} + { {1/ 2}\over  (p+k_1)^2} + {{1/2}\over
(p-k_{2})^2} . \label{eqreductiona}
\end{eqnarray}

\begin{figure}[ht]
\centerline{\includegraphics[height=3cm]{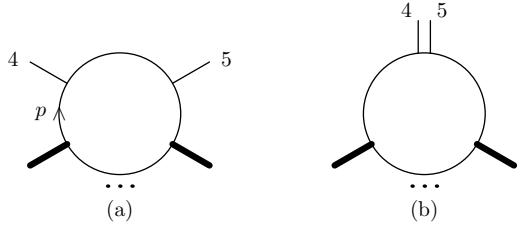} } \caption{A
combination of 2 diagrams with the same adjacent helicity.}
\label{Five}
\end{figure}

For a different choice of reference momenta,
\begin{equation}
\epsilon_4 = \lambda_5\tilde\lambda_4, \qquad \epsilon_5 =
\eta\tilde\lambda_5,
\end{equation}
the tensor reduction of the 2 diagrams shown in Fig.~\ref{Five} is
given as follows:
\begin{eqnarray}
A_{a} + A_{b} & =  & {P_4(p)\, P_5(p-k_4)\over
p^2\,(p-k_4)^2\,(p-k_{45})^2 } +
{P_{45}(p) \over p^2(p-k_{45})^2} \nonumber \\
& = &  - { (\eta\tilde\lambda_4,p)\over p^2\,(p-k_{4})^2} +
{\langle\eta\,5\rangle\over 2 \, \langle4\,5\rangle }\, \left[ {
1\over (p-k_{45})^2} - {1\over p^2} \right] . \label{eqreductionb}
 \end{eqnarray}
Setting $\eta=\lambda_4$, the above equation agrees with
eq.~(\ref{eqreductiona}).

\section{A summary of the rational parts for triangle and box
integrals}

For quick reference we list here the explicit formulas for the
rational parts of some Feynman integrals which are needed to
compute the 5-gluon amplitudes. The derivation can be found in
\cite{xyzi}.

Firstly, for the bubble integral we have:
\begin{eqnarray}
I_2(\epsilon_1,\epsilon_2) & = & \int { {\rm d}^D p \over i
\pi^{D/2} } \, {
(\epsilon_1,p)(\epsilon_2,p) \over p^2 ( p+K)^2} \nonumber \\
& = & {1\over 18}\, ( (\epsilon_1,K)\, (\epsilon_2,K) -  2 \,
K^2\, (\epsilon_1,\epsilon_2) ) .
\end{eqnarray}
where $K$ is the sum of momenta on one side of the bubble diagram.
The above result first appeared in \cite{BDK}.

For 2-mass triangle integral with external momenta
$\{k_1,K_2,K_3\}$ and $k_1^2=0$, the degree 2 integral is:
\begin{eqnarray}
I_3 (\epsilon_1,\epsilon_2) & \equiv &   \int {{\rm d}^D p \over i
\pi^{D/2}} \,
  { (\epsilon_1, p) \, (\epsilon_2 , p) \,  \over p^2(p-k_1)^2
  (p+K_3)^2}    \nonumber \\
 & = &      {1\over 2}\, (\epsilon_1, \epsilon_2) +
   {(K^2_2 + K^2_3)\over 2 (K^2_2-K^2_3)^2 }
  \, (\epsilon_1, k_1) \, (\epsilon_2, k_1)  \nonumber \\
& + & {( (\epsilon_1, K_2)\,(\epsilon_2,k_1)  -(\epsilon_1, k_1)\,
(\epsilon_2, K_3) ) \over 2(K^2_2-K^2_3)}
 , \label{twomassone} \\
I_3 (\epsilon_1,\epsilon_2) & =  &
     {1\over2}\, (\epsilon_1, \epsilon_2) +  { (\epsilon_1, K_2) \,
(\epsilon_2, k_1)  \over 2(K_2^2-K^2_3)}  , \qquad (\epsilon_1,
k_1) = 0, \label{twomasstwo}
\end{eqnarray}
and the degree 3 integral is:
\begin{eqnarray}
I_3 (\epsilon_i) & \equiv &  \int {{\rm d}^D p \over i \pi^{D/2}}
\,\,
  { (\epsilon_1, p)  \, (\epsilon_2 ,  p-k_1) \, (\epsilon_3, p) \,
 \over p^2\, (p-k_1)^2 \,   (p+K_3)^2}
  \nonumber \\
  &   =  &   {1\over 36} \Big( (\epsilon_2,  4\,K_2 -7\, k_1)\,(\epsilon_1,
\epsilon_3) -(2 \leftrightarrow 3)  + 4 (\epsilon_1, K_2)\,
(\epsilon_2, \epsilon_3) \Big)  \nonumber \\
& - & {(K^2_2 + K^2_3)  \over 6\,(K^2_2-K^2_3)^2 } \, (\epsilon_1,
K_2) \,   (\epsilon_2, k_1) \, (\epsilon_3, k_1)
\nonumber \\
 & - & {(\epsilon_1, K_2)\, ((\epsilon_2, k_1)\,
(\epsilon_3, K_3) - (\epsilon_2 , K_2)\, (\epsilon_3, k_1))\over
6\, (K^2_2-K^2_3)}
\nonumber \\
& - & { (K^2_2 + K^2_3) \over 12\, (K^2_2-K^2_3) } \, (
(\epsilon_1, \epsilon_2)\, (\epsilon_3, k_1) + (\epsilon_1 ,
\epsilon_3)\, (\epsilon_2, k_1) )  , \label{twomassthree}
\end{eqnarray}
where $\epsilon_1$ satisfies the physical condition
$(\epsilon_1,k_1)$ = 0 and $\epsilon_{2,3}$ are arbitrary
4-dimensional polarization vectors.

For five point cases we need only 1-mass box
integrals. The 1-mass box integrals can be obtained simply by
setting the mass of one massive external line to 0 from the 2-mass-easy
 box integrals. So we list the formulas of two-mass-easy box integrals here.
 The four external momenta of the 2-mass-easy
box diagram are denoted as $\{k_1, K_2,k_3,K_4\}$.

For degree 3 two-mass-easy box integrals, we have:
\begin{eqnarray}
I_4^{2me} (\lambda_3\tilde\lambda_1,\epsilon_2, \epsilon_3)  =
 {\langle 3 |K_2|1 ]   \over 2} \, \left[ { (\epsilon_2 , k_3) \,
(\epsilon_3 , k_3) \over (K_2^2 - t) (K_4^2 - s) } - (k_3
\rightarrow k_1, s \leftrightarrow t) \right] ,
\end{eqnarray}
where $s=(k_1+K_2)^2$ and $t=(K_2+k_3)^2$. If two of the
polarization vectors satisfy the physical condition, i.e.
$(\epsilon_1,k_1)=0$ and $(\epsilon_3,k_3)=0$, we have
\begin{eqnarray}
I_4^{2me} (\epsilon_1,\epsilon_2, \epsilon_3) & = &
 - {(\epsilon_1,k_3)\,(\epsilon_3,k_1)\over 2\, (k_1,k_3) } \,\left[
 {(\epsilon_2,k_3)\over K_2^2 - t } + {(\epsilon_2,k_1)\over K_4^2 -t} \right]
\nonumber \\
&   & \hskip -1cm - { (\epsilon_1,K_2)
\,(\epsilon_2,k_1)\,(\epsilon_3,k_1) \over 2\, (K_2^2 - s)\,(K_4^2
- t)} - { (\epsilon_1,k_3) \,(\epsilon_2,k_3)\,(\epsilon_3,K_4)
\over 2\, (K_2^2 - t)\,(K_4^2 - s)} .
\end{eqnarray}

In order to give the formulas for the rational parts of degree
4 polynomials, we define:
\begin{equation}
I_4 (\epsilon_1,\epsilon_2, \epsilon_3, \epsilon_4)   \equiv
 \int { {\rm d}^D p \over i \pi^{D/2}}
\, { (\epsilon_1, p) \, (\epsilon_2 ,  p-k_1) \, (\epsilon_3 ,  p
- K_{12})\, (\epsilon_4, p+K_4)  \over p^2(p-k_1)^2 (p-K_{12})^2
(p+K_4)^2}  ,
\end{equation}
where $K_{12} = k_1 + K_2$. For the two-mass-easy case, we have
\begin{eqnarray}
 I_4(\lambda_3\tilde\lambda_1, \epsilon_2,\lambda_1\tilde\lambda_3,
 \epsilon_4)  & = &  -  {1\over 4} \left( {K_2^2 + s \over K_2^2 - s } +
 {K_4^2 + t \over K_4^2 - t }\right) \,(\epsilon_2,k_1)(\epsilon_4,k_1)
 \nonumber \\
&   & \hskip -1.5cm
 - {1\over 4} \left( {K_2^2 + t \over K_2^2 - t
} +  {K_4^2 + s \over K_4^2 - s }\right)
\,(\epsilon_2,k_3)(\epsilon_4,k_3)
-  {5\over 9} \, (k_1,k_3)( \epsilon_2,  \epsilon_4) \nonumber \\
& +  &  {4\over 9}\, \Big( (\epsilon_2,k_1)(\epsilon_4,k_3) +
 (\epsilon_2,k_3)(\epsilon_4,k_1) \Big)   , \\
I_4(\lambda_1\tilde\lambda_3, \epsilon_2,\lambda_1\tilde\lambda_3,
 \epsilon_4)  & = &  {5\over 9}\, \langle 1|\epsilon_2|3] \, \langle1|\epsilon_4|3]
 \nonumber \\
 &  & \hskip -2cm + {\langle1|K_2|3]^2 \over 3}\, \left[
 {(\epsilon_2,k_1)\,(\epsilon_4,k_1) \over (K_2^2 - s) \, (K_4^2 - t)} +
 {(\epsilon_2,k_3)\,(\epsilon_4,k_3) \over (K_2^2 - t) \, (K_4^2 - s)} \right]   .
\end{eqnarray}
Other cases can be either obtained by conjugation or by relabelling
$k_{1,3}$.

The 3-mass triangle and 2-mass-hard box integrals are not needed
for the computation of the 5-gluon amplitudes. The rational parts
for these integrals can be found in \cite{xyzi,xyziii}.

\section{MHV: $R(1^-2^-3^+4^+5^+)$}
Now we begin the computation of the rational parts of the 5-gluon
amplitudes. The reader should consult eq. (31) in \cite{xyzi}
for the detailed definition of the rational part.
In this section we compute the rational part for the
helicity configuration $(1^-2^-3^+4^+5^+)$. For this helicity
configuration we choose the following polarization vectors:
\begin{eqnarray}
\epsilon_1 = {\lambda_1\tilde\lambda_2\over [1\,2] } , & &
\epsilon_2 = {\lambda_2\tilde\lambda_1\over [2\,1] } , \\
\epsilon_3 = {\lambda_4\tilde\lambda_3\over \langle4\,3\rangle
 } , & &
\epsilon_5 = {\lambda_4\tilde\lambda_5\over \langle4\,5\rangle
 } , \qquad
\epsilon_4 = {\eta\tilde\lambda_4\over \langle\eta\,4\rangle } .
\end{eqnarray}
In order to keep the symmetry under $1\leftrightarrow2$,
$3\leftrightarrow5$ manifest, we leave the reference momentum for
$\epsilon_4$ arbitrary, i.e. $\eta$ is an arbitrary (holomorphic)
spinor. The final result should be independent of $\eta$. In the
following formulas for $R_i$'s, we will omit all the denominators of
the polarization vectors.

\begin{figure}[ht]
\centerline{\includegraphics[height=3cm]{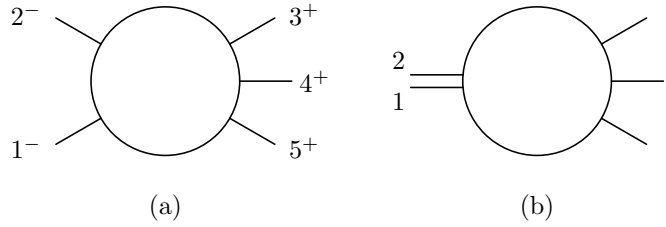} }
\caption{The first diagram is the only 5-point diagram. Its
combination with the pinched $k_{1,2} \to k_{12}$ 4-point diagram
leads to triangle diagrams only by making use of the reduction
formula (\ref{eqreductiona}).} \label{FirstSet}
\end{figure}

\begin{figure}[ht]
\centerline{\includegraphics[height=6cm]{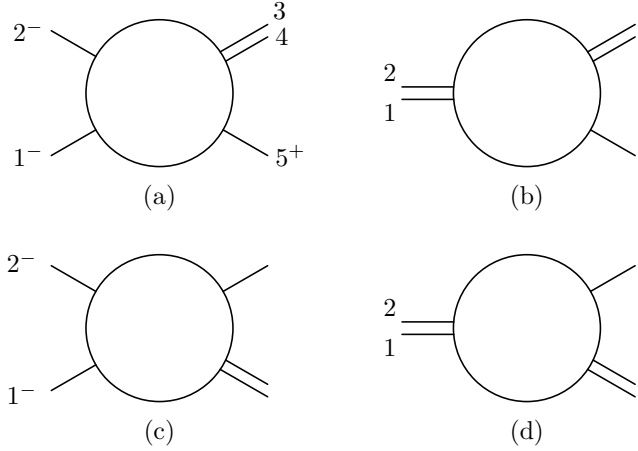} }
\caption{These 4 Feynman diagrams are reduced as the 2 diagrams in
Fig.~\ref{FirstSet}.} \label{SecondSet}
\end{figure}

\begin{figure}[ht]
\centerline{\includegraphics[height=6cm]{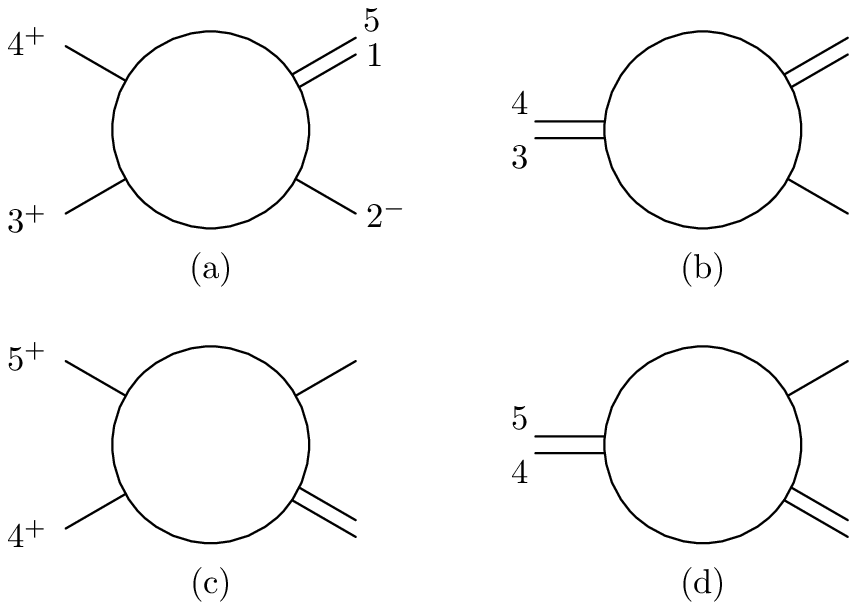} }
\caption{The 3rd set of Feynman diagrams. They can also be reduced
easily because of the same adjacent helicity $3^+4^+$ or
$4^+5^+$.} \label{ThirdSet}
\end{figure}

\begin{figure}[ht]
\centerline{\includegraphics[height=3cm]{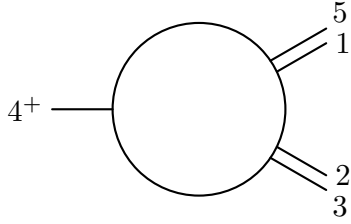} }
\caption{The only 2-mass triangle diagram which does not combine
with a higher point diagram. } \label{FourthSet}
\end{figure}

\begin{figure}[ht]
\centerline{\includegraphics[height=3cm]{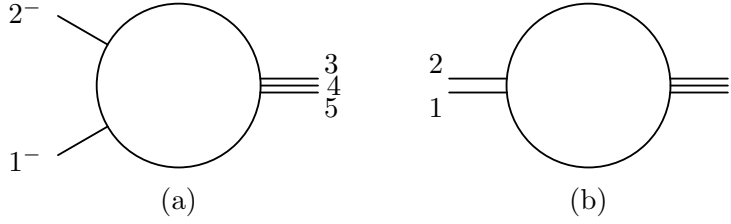} }
\caption{This set has 2 Feynman diagrams. Their reduction leads to
tadpole (one-point) diagrams which are zero in dimensional
regularization.} \label{FifthSet}
\end{figure}

\begin{figure}[ht]
\centerline{\includegraphics[height=4cm]{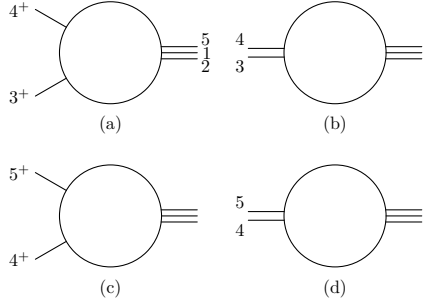} } \caption{
This set has 4 Feynman diagrams. Tensor reduction by making use of
eq.~(\ref{eqreductionb}) leads to bubble diagrams.}
\label{SixthSet}
\end{figure}

\begin{figure}[ht]
\centerline{\includegraphics[height=4cm]{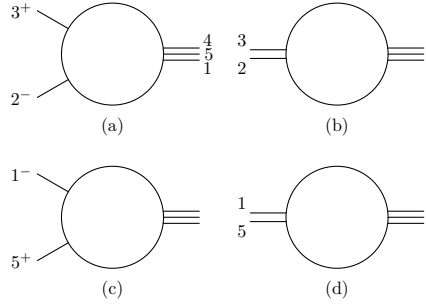} } \caption{
The last set has two pairs of Feynman diagrams with different
adjacent helicities. Each combination is identically zero. }
\label{SeventhSet}
\end{figure}

We classify all the 21 diagrams into  7 sets. They are shown in
Figs.~\ref{FirstSet} to ~\ref{SeventhSet}. Now let us present the
results of the rational parts for these 7 sets of Feynman diagrams.

For the two Feynman diagrams given in Fig.~\ref{FirstSet}, by
using the tensor reduction formula given in
eq.~(\ref{eqreductiona}), they are reduced to just triangle
diagrams. Because the other three external lines have the same
helicity, these triangle diagrams can be further reduced to
bubble diagrams which can be computed easily. We note that some
extra terms must be added for tensor reduction of the box or
triangle integrals \cite{xyzi}. The final result for the rational
part is exceptionally simple and is given as follows:
\begin{eqnarray}
R_1 =  -{1\over 18}\, ((\epsilon_4,\epsilon_5)\epsilon_3 +
(\epsilon_3,\epsilon_4)\epsilon_5 , k_1-k_2) .
\end{eqnarray}

For the four Feynman diagrams shown in Fig.~(\ref{SecondSet}), we
can use the same tensor reduction formula as above.  Because we
use the tensor reduction formula for the box integral, we should add
an extra term. The rational part of these four diagrams is:
\begin{eqnarray}
R_2 &  = &   {1\over 18}\, ((\epsilon_{34},k_2)(\epsilon_5,k_1) +
(\epsilon_{3},k_2)(\epsilon_{45},k_1)) \nonumber \\
& + & {1\over 18}\, (2s_{51}-s_{34}-3s_{12})\,
(\epsilon_{34},\epsilon_5) \nonumber \\
& + & {1\over 18}\, (2s_{23}-s_{45}-3s_{12})\,
(\epsilon_{3},\epsilon_{45}) .
\end{eqnarray}

The four diagrams shown in Fig.~\ref{ThirdSet} can also be reduced
simply. Because we leave the reference momentum of $\epsilon_4$
arbitrary, the reduction is just to triangle integrals by using
eq.~(\ref{eqreductionb}). The rational part is:
\begin{eqnarray}
R_3 & = & - {1\over9}\, s_{23}(\tilde\epsilon_{23},\epsilon_{51}) -
{\langle\eta\,4\rangle \over18\,\langle3\,4\rangle}\,
s_{51}(\epsilon_2, \epsilon_{51}) - {1\over6}\,
(\eta\tilde\lambda_3,k_4)\, (\epsilon_2,\epsilon_{51})
\nonumber \\
& + & {1\over9}\, s_{51}(\tilde\epsilon_{51},\epsilon_{23}) -
{\langle\eta\,4\rangle \over18\,\langle5\,4\rangle}\, s_{23}(\epsilon_1,
\epsilon_{23}) - {1\over6}\, (\eta\tilde\lambda_5,k_4)\,
(\epsilon_1,\epsilon_{23}),
\end{eqnarray}
where $\tilde\epsilon_{23} = \epsilon_{23}|_{\epsilon_3 \to
\eta\tilde\lambda_3}$ and  $\tilde\epsilon_{51} =
\epsilon_{51}|_{\epsilon_5 \to \eta\tilde\lambda_5}$, i.e.:
\begin{eqnarray}
\tilde\epsilon_{23} & = & {[1\,3]\over [2\,3] }\,
\eta\tilde\lambda_3 - {\langle 2 \,\eta\rangle \over  \langle 2
\,3\rangle} \, \lambda_2\tilde\lambda_1 + { \langle 2
\,\eta\rangle  \, [3\, 1] \over 2\, \langle 2 \,3\rangle \, [3\,2]
} \, (k_2-k_3), \\
\tilde\epsilon_{51} & = &  {\langle 1 \,\eta\rangle \over \langle
1 \,5\rangle }\, \lambda_1\tilde\lambda_2 - {[2\,5]\over [1\,5]}
\, \eta\tilde\lambda_5 + { \langle 1 \,\eta\rangle \, [5\,2] \over
2\, \langle 1\,5\rangle \, [5\, 1] }\, (k_5-k_1).
\end{eqnarray}

The Feynman diagram shown in Fig.~\ref{FourthSet} is the only 2-mass
 triangle diagram which does not combine with a higher point
diagram. This actually gives rise to spurious pole terms which can
only be cancelled by the contribution from cut-constructible part.
The rational part is computed by using eq.~(\ref{twomassthree}):
\begin{eqnarray}
R_4 &  = &  {1\over 36} \, ( 7 (\epsilon_4,  \epsilon_{51})
(\epsilon_{23}, k_{4}) -
7 (\epsilon_4,  \epsilon_{23}) (\epsilon_{51}, k_{4}) +
4 (   \epsilon_{23}, \epsilon_{51})(\epsilon_{4}, k_{51}) )
\nonumber \\
& - &  {s_{51} +  s_{23} \over 12(s_{51} -  s_{23})  }
\, ( (\epsilon_4,  \epsilon_{51}) (\epsilon_{23}, k_{4}) +
(\epsilon_4,  \epsilon_{23}) (\epsilon_{51}, k_{4}) )
\nonumber \\
& - &    {s_{51} +  s_{23} \over 6(s_{51} - s_{23})^2 } \,
(\epsilon_4, k_{51})(\epsilon_{51}, k_{4})(\epsilon_{23}, k_{4})
\nonumber \\
& - &  {1\over 4} ( \epsilon_4,
(\epsilon_5,\epsilon_{1})\epsilon_{23} +
(\epsilon_2,\epsilon_3)\epsilon_{51})
\nonumber \\
& - & {1\over4}\, { ( \epsilon_4, k_{51})\,
((\epsilon_5,\epsilon_{1})\epsilon_{23} +
(\epsilon_2,\epsilon_3)\epsilon_{51},k_4) \over s_{51}-s_{23}} .
\end{eqnarray}

The 2 Feynman diagrams shown in Fig.~\ref{FifthSet} are reduced to
tadpole (one-point) diagrams which are zero in dimensional
regularization. The only contribution is from the extra terms and
the result is:
\begin{eqnarray}
R_5 & = & -{1\over6}\, s_{12}(\epsilon_{345},k_1) + {1\over4}
\, s_{12}\, ( (\epsilon_3,\epsilon_{45}) + (\epsilon_{34},\epsilon_5)).
\end{eqnarray}

The 4 Feynman diagrams shown in Fig.~\ref{SixthSet} are reduced to
bubble diagrams by making use of eq.~(\ref{eqreductionb}) because
the 2 massless external lines $k_{3,4}$ or $k_{4,5}$ have the same
adjacent helicity. The rational part is:
\begin{eqnarray}
R_6 & = & (\eta\tilde\lambda_3,k_4)\, \left[
 {1\over4} \, ( (\epsilon_5,\epsilon_{12}) + (\epsilon_{51},\epsilon_2))
 + {1\over6}\,  (\epsilon_{512},k_4) \right] \nonumber \\
& + & (\eta\tilde\lambda_5,k_4)\, \left[
 {1\over4} \, ( (\epsilon_1,\epsilon_{23}) + (\epsilon_{12},\epsilon_3))
- {1\over6}\,  (\epsilon_{123},k_4) \right]  .
\end{eqnarray}

The last 4 diagrams shown in Fig.~\ref{SeventhSet} are two pairs
of Feynman diagrams with different adjacent helicities. By
explicit computation we found that  each combination is
identically zero. This result is true for more general cases where
the composite polarization  vector can be arbitrary.

Having computed all the 21 diagrams separately, the complete
rational part is obtained by adding them together. We have
\begin{eqnarray}
R & = & {1\over [1\,2]^2\,\langle3\, 4\rangle \, \langle4\,
5\rangle \, \langle\eta\,4\rangle }\, \left[
 -{1\over 18}\, ((\epsilon_4,\epsilon_5)\epsilon_3 +
(\epsilon_3,\epsilon_4)\epsilon_5 , k_1-k_2) \right. \nonumber \\
& + & {1\over 18}\, ((\epsilon_{34},k_2)(\epsilon_5,k_1) +
(\epsilon_{3},k_2)(\epsilon_{45},k_1)) \nonumber \\
& + & {1\over 18}\, (2s_{51}-s_{34}-3s_{12})\,
(\epsilon_{34},\epsilon_5) +  {1\over 18}\,
(2s_{23}-s_{45}-3s_{12})\, (\epsilon_{3},\epsilon_{45}) \nonumber
\\ & - & {1\over9}\, s_{23}(\tilde\epsilon_{23},\epsilon_{51}) -
{\langle\eta\,4\rangle \over18\,\langle3\,4\rangle}\,
s_{51}(\epsilon_2, \epsilon_{51}) - {1\over6}\,
(\eta\tilde\lambda_3,k_4)\, (\epsilon_2,\epsilon_{51})
\nonumber \\
& + & {1\over9}\, s_{51}(\tilde\epsilon_{51},\epsilon_{23}) -
{\langle\eta\,4\rangle \over18\,\langle5\,4\rangle}\,
s_{23}(\epsilon_1, \epsilon_{23}) - {1\over6}\,
(\eta\tilde\lambda_5,k_4)\, (\epsilon_1,\epsilon_{23}) \nonumber
\\ & - & {1\over6}\, s_{12}(\epsilon_{345},k_1) + {1\over4} \, s_{12}\, (
(\epsilon_3,\epsilon_{45}) + (\epsilon_{34},\epsilon_5)) \nonumber
\\ & + & (\eta\tilde\lambda_3,k_4)\, \left[
 {1\over4} \, ( (\epsilon_5,\epsilon_{12}) + (\epsilon_{51},\epsilon_2))
 + {1\over6}\,  (\epsilon_{512},k_4) \right] \nonumber \\
& + & (\eta\tilde\lambda_5,k_4)\, \left[
 {1\over4} \, ( (\epsilon_1,\epsilon_{23}) + (\epsilon_{12},\epsilon_3))
- {1\over6}\,  (\epsilon_{123},k_4) \right]  \nonumber \\ & + &
{1\over 36} \, ( 7 (\epsilon_4,  \epsilon_{51}) (\epsilon_{23},
k_{4}) - 7 (\epsilon_4,  \epsilon_{23}) (\epsilon_{51}, k_{4}) + 4
( \epsilon_{23}, \epsilon_{51})(\epsilon_{4}, k_{51}) )
\nonumber \\
& - &  {s_{51} +  s_{23} \over 12(s_{51} -  s_{23})  } \, (
(\epsilon_4,  \epsilon_{51}) (\epsilon_{23}, k_{4}) + (\epsilon_4,
\epsilon_{23}) (\epsilon_{51}, k_{4}) )
\nonumber \\
& - &    {s_{51} +  s_{23} \over 6(s_{51} - s_{23})^2 } \,
(\epsilon_4, k_{51})(\epsilon_{51}, k_{4})(\epsilon_{23}, k_{4})
\nonumber \\
& - &  {1\over 4} ( \epsilon_4,
(\epsilon_5,\epsilon_{1})\epsilon_{23} +
(\epsilon_2,\epsilon_3)\epsilon_{51})
\nonumber \\
& - & \left. {1\over4}\, { ( \epsilon_4, k_{51})\,
((\epsilon_5,\epsilon_{1})\epsilon_{23} +
(\epsilon_2,\epsilon_3)\epsilon_{51},k_4) \over s_{51}-s_{23}}
\right] .\label{MHV1}
\end{eqnarray}
For the same helicity configuration, the rational part of the amplitude obtained by
Bern, Dixon and Kosower as given in \cite{BDK} is :
\begin{eqnarray}
\tilde R & = & {2 \over 9} \, {\langle 1 \, 2\rangle^4 \over
\langle 1 \, 2\rangle \, \langle 2 \, 3\rangle \, \langle3 \,
4\rangle \,\langle 4 \, 5\rangle \, \langle 5 \, 1 \rangle} -{1
\over 3} \, {\langle 3 \, 5\rangle \, [3\,5]^3 \over [1\,2] \,
[2\,3] \, \langle 3 \, 4\rangle \, \langle 4 \, 5\rangle \,
[5\,1]} \nonumber \\ & + & {1 \over 3} \, {\langle 1 \, 2\rangle
\, [3\,5]^2 \over [2\,3] \, \langle 3 \, 4\rangle \, \langle 4 \,
5\rangle \, [5\,1]} + {1 \over 6} \, {\langle 1 \, 2\rangle \,
[3\,4] \, \langle 4 \, 1\rangle \, \langle 2 \, 4\rangle \, [4\,5]
\over s_{23} \, \langle 3 \, 4\rangle \, \langle 4 \, 5\rangle \,
s_{51} } \nonumber \\ & - & {1 \over 6} \, {[3\,4] \, \langle 4 \,
1\rangle \, \langle 2 \, 4\rangle \, [4\,5] \, (\langle 2 \,
3\rangle \, [3\,4] \, \langle 4 \, 1\rangle + \langle 2 \,
4\rangle \, [4\,5] \, \langle 5 \, 1\rangle) \over s_{23} \,
\langle 3 \,
4\rangle \, \langle 4 \, 5\rangle \, s_{51} } \nonumber \\
&& \times \, {s_{51} + s_{23} \over (s_{51} - s_{23})^2 }\, ,\label{MHV1Bern}
\end{eqnarray}
which can be obtained from the full amplitude
eq. (8) and (10) in \cite{BDK} by making the replacement
\begin{eqnarray}
  \mathrm{L}_0(r)&\rightarrow&0\,, \nonumber\\
  \mathrm{L}_1(r)&\rightarrow&{1\over 1-r}\, , \nonumber\\
  \mathrm{L}_2(r)&\rightarrow&\frac 1 2 {(r+1)\over r(1-r)^2}\,
\end{eqnarray}
and discarding $V^f$ which is absorbed in the cut constructible part.
One can find detailed discussions about how to construct the rational
part from Bern etc.'s results in  \cite{BDKB,BDKC}.
We did not find an easy way to prove that these results ($R$ and
$\tilde R$) agree with each other. With the help of Mathematica
one can easily check the following result:
\begin{equation}
R = \, {1 \over 2} \, \tilde R.
\end{equation}
This shows that the rational part computed directly from Feynman
integrals agrees with the well-known result of Bern, Dixon and
Kosower \cite{BDK}.
The factor $1/2$ is because $R$ is for the amplitude with a
real scalar circulating in the loop according to our definition in
eq. (31) in \cite{xyzi}, while in their notation
 $\tilde{R}$ is for the amplitude with a complex scalar.

\section{MHV: $R(1^-2^+3^-4^+5^+)$}

For this helicity configuration we choose the following
polarization vectors:
\begin{eqnarray}
\epsilon_1 = {\lambda_1\tilde\lambda_5\over [1\,5] } , & &
\epsilon_3 = {\lambda_3\tilde\lambda_4\over [3\,4] } , \\
\epsilon_4 = {\lambda_5\tilde\lambda_4\over \langle5\,4\rangle
 } , & &
\epsilon_5 = {\lambda_4\tilde\lambda_5\over \langle4\,5\rangle
 } , \qquad
\epsilon_2 = {\eta\tilde\lambda_2\over \langle\eta\,2\rangle } .
\end{eqnarray}
The reference momentum of $\epsilon_2$ is arbitrary. As before, in
the following formulas for $R_i$'s, we will omit all the
denominators of the polarization vectors.

\begin{figure}[ht]
\centerline{\includegraphics[height=3cm]{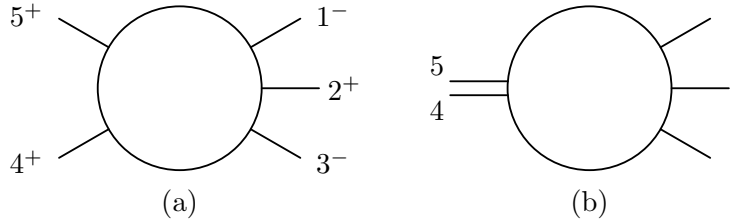}} \caption{6
Feynman diagrams with the same adjacent helicity. } \label{MHVba}
\end{figure}

\begin{figure}[ht]
\centerline{\includegraphics[height=4.5cm]{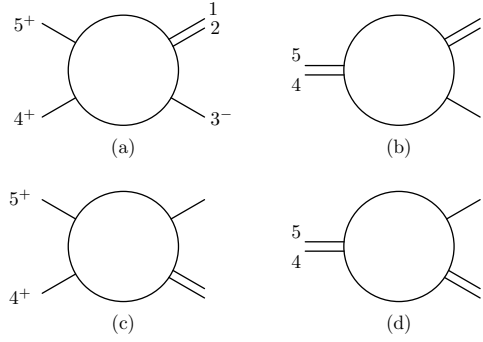}}
\caption{This set consists of 4  Feynman diagrams which can be
reduced easily because of the same adjacent helicity.}
\label{MHVbb}
\end{figure}

\begin{figure}[ht]
\centerline{\includegraphics[height=3cm]{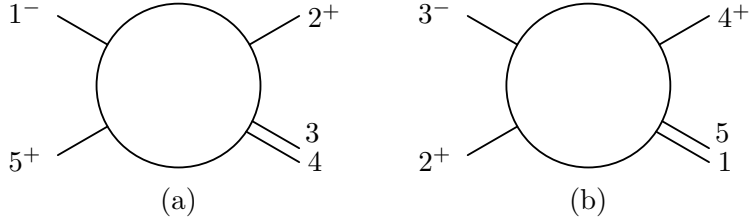}} \caption{2
Feynman diagrams which are computed by using the 2-mass-easy
formulas as given in Sect.~4.} \label{MHVbc}
\end{figure}

\begin{figure}[ht]
\centerline{\includegraphics[height=2.5cm]{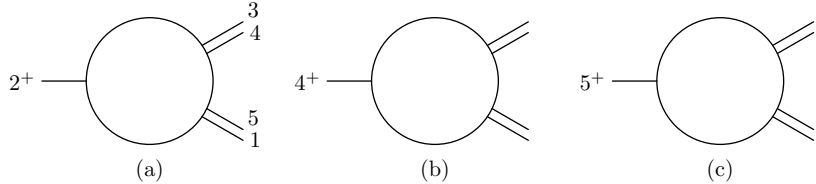}}
\caption{The 4th set consists of 3 two-mass triangle diagrams.}
\label{MHVbd}
\end{figure}

\begin{figure}[ht]
\centerline{\includegraphics[height=3cm]{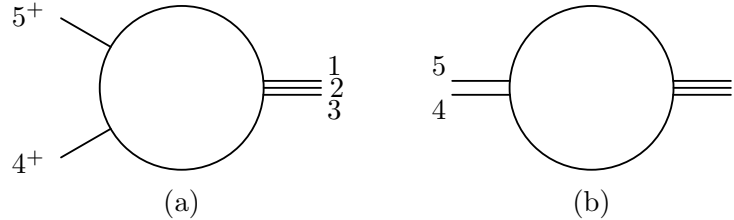}} \caption{2
Feynman diagrams with the same adjacent helicity $k_{4,5}$. }
\label{MHVbe}
\end{figure}

The 21 Feynman diagrams are classified into 6 sets. Apart from the
last set, they are shown in Figs.~\ref{MHVba} to \ref{MHVbe}.
Let us compute the rational part from each set in turn.

The 1st set consists of 2 Feynman diagrams which include the only
pentagon diagram. Tensor reduction is easy by using
eq.~(\ref{eqreductiona}) which gives rise to triangle diagrams.
The rational parts of the reduced integrals can be computed by
using eqs.~(\ref{twomassone}) to (\ref{twomassthree}) and we have:
\begin{eqnarray}
R_{1}  &  = & - {1\over 18} \, (
(\epsilon_2,\epsilon_3)\epsilon_1 +
(\epsilon_1,\epsilon_2)\epsilon_3 +
(\epsilon_1,\epsilon_3)\epsilon_2, k_4-k_5) \nonumber \\
& + & {1\over 12} \, (\epsilon_1,\epsilon_2)(\epsilon_3, k_2 -
k_{45})  - {1\over 12} \,   (\epsilon_2,\epsilon_3)(\epsilon_1,
k_2 - k_{45}) \nonumber \\
& + & {s_{34}+ s_{51} \over 12( s_{34} - s_{51} )} \, (    (
(\epsilon_2,\epsilon_3)\epsilon_1 +
(\epsilon_2,\epsilon_1)\epsilon_3, k_2) \nonumber \\
& + &   {s_{34}+ s_{51} \over 6( s_{34} - s_{51} )^2} \,
(\epsilon_2,k_{34})(\epsilon_3, k_2)(\epsilon_1,k_2)  .
\end{eqnarray}

The 2nd set of Feynman diagrams consist of 4 diagrams. They are
reduced identically as in the above. We only need to add an extra
term because of the box reduction in $D=4$. The rational part from
set 2 is:
\begin{eqnarray}
R_{2}  &  = & \left(  -{1\over6}\,  s_{45} +  {1\over9}\, s_{51}
- {1\over 18} \,  s_{23}  \right) \, (\epsilon_1,\epsilon_{23})
\nonumber \\
&  + &  \left(  -{1\over6}\,  s_{45} +  {1\over9}\, s_{34} -
{1\over 18}
\, s_{12}  \right) \, (\epsilon_{12},\epsilon_{3}) \nonumber \\
& + & {1\over 18} \, ( (\epsilon_{23},k_4)(\epsilon_1,k_5) +
(\epsilon_{3},k_4)(\epsilon_{12},k_5) ) .
\end{eqnarray}

The two diagrams shown in Fig.~\ref{MHVbc} are the most
complicated for the 5-gluon amplitude. There is no simple
reduction formula to simplify the computation. One needs to use
the 2-mass-easy formulas as given in Sect.~4. The explicit result
of the rational part from the second diagram is:
\begin{eqnarray}
R_3{(2)} & = & {\langle \eta\, 4\rangle\, \langle 5\, 2\rangle
\over 18 \, \langle 2\,4\rangle \, \langle4\,2\rangle } \, \left[
(\epsilon_3,k_2)(\epsilon_{51},k_4) -
2(k_2,k_4)(\epsilon_3,\epsilon_{51})  \right]
\nonumber \\
& + &  {\langle \eta\, 4\rangle \over \langle 2\,4\rangle  } \,
\left[ {1\over9}\, ((\epsilon_4, \epsilon_{51})\epsilon_3+
(\epsilon_3,\epsilon_{51})\epsilon_4,k_2) +   {
(\epsilon_4,k_{51}) \,
(\epsilon_{51},k_4) \, (\epsilon_3,k_2)    \over 6 \,(s_{51}-s_{23}) } \right]
\nonumber \\
& + &  {\langle 5\, 2\rangle \over \langle 4\,2\rangle  } \,
\left[ {1\over12}\, ((\epsilon_2, \epsilon_{3})\, \epsilon_{51} -
(\epsilon_2, \epsilon_{51})\,\epsilon_{3},k_2)  \right. \nonumber \\
& & + {1\over 18}( 2 (\epsilon_3, \epsilon_{51})\epsilon_2 -
(\epsilon_2,\epsilon_3)\epsilon_{51},k_4) \nonumber \\
& & - { s_{34} + s_{51}\over 12 \,(s_{34}-s_{51}) }\, (
(\epsilon_2,\epsilon_3)\,  \epsilon_{51}+
(\epsilon_2,\epsilon_{51})\, \epsilon_3, k_2)
\nonumber \\
&&\left. + { (s_{34}+s_{51})
\over 6 \,(s_{34}-s_{51})^2 }\, (\epsilon_2,k_{51}) \,
(\epsilon_{3},k_2)\, (\epsilon_{51},k_2)
\right]   \nonumber \\
& + & {\langle \eta\, 2\rangle\, \langle 5\, 4\rangle \over
\langle 4\,2\rangle \, \langle2\,4\rangle } \,
\left[ -\frac 5 9 (\epsilon_{3},\epsilon_{51})(k_{2},k_{4})
+\frac 4 9(\epsilon_{3},k_{2})(\epsilon_{51},k_{4})\right.
\nonumber\\ &&  \left.
+\frac 1 4 (\epsilon_{3},k_{2})(\epsilon_{51},k_{2})
\left(1+{s_{34}+s_{51}\over s_{34}-s_{51}}\right) \right]  .
\end{eqnarray}
The rational part from the first diagram in  Fig.~\ref{MHVbc} can
be obtained from the above result by the symmetry transformation:
\begin{equation}
R_3(1)  = - R_3(2)|_{ 1\leftrightarrow3, 4 \leftrightarrow5}.
\end{equation}
We also set $R_3 = R_3(1) + R_3(2)$.

The rational part of the 3 two-mass triangle diagrams shown in
Fig.~\ref{MHVbd} can be directly computed by using
eqs.~(\ref{twomassthree}) and (\ref{twomasstwo}). The rational
part from the first diagram  in Fig.~\ref{MHVbd} is denoted by
$R_4{(0)}$ and is given by:
\begin{eqnarray}
R_4{(0)}  &  = &  {1\over 36} \, ( 7 (\epsilon_2, \epsilon_{34})
(\epsilon_{51}, k_{2}) - 7 (\epsilon_2, \epsilon_{51})
(\epsilon_{34}, k_{2}) + 4 (   \epsilon_{34},
\epsilon_{51})(\epsilon_{2}, k_{34}) )
\nonumber \\
& - &  {s_{34} +  s_{51} \over 12(s_{34} -  s_{51})  } \, (
(\epsilon_2,  \epsilon_{34}) (\epsilon_{51}, k_{2}) + (\epsilon_2,
\epsilon_{51}) (\epsilon_{34}, k_{2}) )
\nonumber \\
& - &    {s_{34} +  s_{51} \over 6(s_{34} -  s_{51})^2 } \,
(\epsilon_2, k_{34})(\epsilon_{34}, k_{2})(\epsilon_{51}, k_{2}) .
\end{eqnarray}
The rational part from the third Feynman diagram in
Fig.~\ref{MHVbd} is denoted by $R_4{(1)}$ and we have:
\begin{eqnarray}
R_4{(1)}  &  = &  {1\over 36} \, ( 7 (\epsilon_5, \epsilon_{12})
(\epsilon_{34}, k_{5}) - 7 (\epsilon_5, \epsilon_{34})
(\epsilon_{12}, k_{5}) + 4 (   \epsilon_{12},
\epsilon_{34})(\epsilon_{5}, k_{12}) )
\nonumber \\
& - &  {s_{12} +  s_{34} \over 12(s_{12} -  s_{34})  } \, (
(\epsilon_5,  \epsilon_{12}) (\epsilon_{34}, k_{5}) + (\epsilon_5,
\epsilon_{34}) (\epsilon_{12}, k_{5}) )
\nonumber \\
& - &    {s_{12} +  s_{34} \over 6(s_{12} -  s_{34})^2 }  \,
(\epsilon_5, k_{12})(\epsilon_{12}, k_{5})(\epsilon_{34}, k_{5})
\nonumber \\
&  - & {1\over 4} (\epsilon_1,\epsilon_2)\left[
(\epsilon_5,\epsilon_{34}) + {(\epsilon_5,k_{12})(\epsilon_{34} ,
k_5) \over s_{12} - s_{34} } \right] .
\end{eqnarray}
The rational part from the second diagram  in Fig.~\ref{MHVbd} is
denoted by $R_4{(2)}$ and it can be obtained from $R_4{(1)}$ by
symmetry operation. In total we have:
\begin{equation}
R_4 = R_4{(0)} + R_4{(1)}  - ( R_4{(1)}|_{ 1\leftrightarrow3, 4
\leftrightarrow5}) .
\end{equation}

The 5th set consists of 2 Feynman diagrams which are reduced to
tadpoles. The rational part is:
\begin{equation}
R_5 = - { 1 \over 6}\, (\epsilon_{123},k_4) \, s_{45} + {1\over 4}
\,  ( (\epsilon_{1},\epsilon_{23}) + (\epsilon_{12},\epsilon_{3})
) \, s_{45}.
\end{equation}

Up to now we have computed 13 Feynman diagrams. The remaining 8 Feynman
diagrams consist of 4 one-mass triangle diagrams and 4 bubble
diagrams. As we showed in last section, they are
identically zero for each pair of triangle and bubble diagrams.

The final result for the rational part is:
\begin{eqnarray}
R &  = &  - \, {1\over [1\,5] \, [3\,4] \, \langle 4 \, 5\rangle^2
\, \langle \eta \, 2\rangle  } \, \sum_{i=1}^5 R_i.
\end{eqnarray}
In comparison, the result obtained by Bern, Dixon and Kosower
\cite{BDK} from string theory is:
\begin{eqnarray}
\tilde R & = & {2 \over 9} \, {\langle 1 \, 3\rangle^4 \over
\langle 1 \, 2\rangle \, \langle 2 \, 3\rangle \, \langle3 \,
4\rangle \,\langle 4 \, 5\rangle \, \langle 5 \, 1 \rangle} + {1
\over 3} \, { [2\,4]^2 \, [2\,5]^2 \over [1\,2] \, [2\,3] \,
[3\,4] \, \langle 4 \, 5\rangle \, [5\,1]} \nonumber \\ & - & {1
\over 3} \, {\langle 1 \, 2\rangle \, \langle 4 \, 1\rangle^2 \,
[2\,4]^3 \over \langle 4 \, 5\rangle \, \langle 5 \, 1\rangle \,
\langle 2 \, 4\rangle \, [2\,3] \, [3\,4] \, s_{51} } + {1 \over
3} \, {\langle 3 \, 2\rangle \, \langle 5 \, 3\rangle^2 \,
[2\,5]^3 \over \langle 5 \, 4\rangle \, \langle 4 \, 3\rangle \,
\langle 2 \, 5\rangle \, [2\,1] \, [1\,5] \, s_{34} } \nonumber \\
& + & {1 \over 6} \, {\langle 1 \, 3\rangle^2 \, [2\,4] \, [2\,5]
\over s_{34} \, \langle 4 \, 5\rangle \, s_{51} } - {\langle 1 \,
2\rangle \, \langle 2 \, 3\rangle \, \langle 3 \, 4\rangle \,
\langle 4 \, 1\rangle^2 \, [2\,4]^2 \over \langle 4 \, 5\rangle \,
\langle 5 \, 1\rangle \, \langle 2 \, 4\rangle^2 \, s_{51}} \,
\left[{1 \over s_{51} - s_{23} }-{1 \over s_{34} - s_{51} }\right]
\nonumber \\
& + & {\langle 3 \, 2\rangle \, \langle 2 \, 1\rangle \, \langle 1
\, 5\rangle \, \langle 5 \, 3\rangle^2 \, [2\,5]^2 \over \langle 5
\, 4\rangle \, \langle 4 \, 3\rangle \, \langle 2 \, 5\rangle^2 \,
s_{34}} \, \left[{1 \over s_{34} - s_{51} }-{1 \over s_{12} -
s_{34} }\right]
\nonumber \\
& + & {1 \over 3} \, {\langle 2 \, 3\rangle^2 \, \langle 4 \,
1\rangle^3 \, [2\,4]^3 \over \langle 4 \, 5\rangle \, \langle 5 \,
1\rangle \, \langle 2 \, 4\rangle \, s_{23} \,  s_{51}} {s_{51} +
s_{23} \over (s_{51} -  s_{23})^2 }
\nonumber \\
& - & {1 \over 3} \, {\langle 2 \, 1\rangle^2 \, \langle 5 \,
3\rangle^3 \, [2\,5]^3 \over \langle 5\, 4\rangle \, \langle 4 \,
3\rangle \, \langle 2 \, 5\rangle \, s_{12} \,  s_{34}} {s_{12} +
s_{34} \over (s_{12} -  s_{34})^2 }
\nonumber \\
& + & \left[ {1 \over 6} \, {\langle 1 \, 3\rangle \, [2\,4] \,
[2\,5] \, (\langle 1 \, 5\rangle \, [5\,2] \, \langle 2 \,
3\rangle - \langle 3 \, 4\rangle \, [4\,2] \, \langle 2 \,
1\rangle) \over \langle 4 \, 5\rangle} \right. \nonumber \\ &&
\left. + {1 \over 3} \, {\langle 1 \, 2\rangle^2 \, \langle 3 \,
4\rangle^2 \, \langle 4 \, 1\rangle \, [2\,4]^3 \over \langle 4 \,
5\rangle \, \langle 5 \, 1\rangle \, \langle 2 \, 4\rangle} - {1
\over 3} \, {\langle 3 \, 2\rangle^2 \, \langle 1 \, 5\rangle^2 \,
\langle 5 \, 3\rangle \, [2\,5]^3 \over \langle 5\, 4\rangle \,
\langle 4 \, 3\rangle \, \langle 2 \, 5\rangle} \right] \nonumber
\\ && \times \, {s_{34} + s_{51} \over s_{34} \, s_{51} \, (s_{34}
- s_{51})^2 }.
\end{eqnarray}
As in the other MHV case, we did not find an easy way to prove that
these results ($R$ and $\tilde R$) agree with each other. With the
help of Mathematica one can easily check the following result:
\begin{equation}
R = \, {1\over 2} \, \tilde R.
\end{equation}
This shows that the rational part computed directly from Feynman
integrals agrees with the well-known result of Bern, Dixon and
Kosower \cite{BDK}.

\section{Conclusion}

In this paper, by using the formalism developed in the previous paper \cite{xyzi},
we computed explicitly the rational parts of the 5-gluon one-loop amplitudes for the
2 MHV helicity configurations.
The results are in agreement with the well-known results of Bern, Dixon and  Kosower
\cite{BDK}.  Intermediate results for some
combinations of Feynman diagrams are presented explicitly. The explicit result for
the MHV
configuration ($--+++$) as given in eq.~(\ref{MHV1}) is more complicated than the previous
known result eq.~(\ref{MHV1Bern}). In fact the use of the composite polarization vectors
hides some
more complexities of our results.
The correctness and usefulness of the formalism as developed in \cite{xyzi} will be
further strengthed in \cite{xyziii} by
computing the rational parts of the 6-gluon one-loop amplitudes for all possible
helicity configurations.

\section*{Acknowledgments}

CJZ would like to thank J. -P. Ma for constant encouragements,
helpful discussions and careful reading of the paper. His
(financial) support (to buy a computer which was still in use
today) actually goes back to the much earlier difficult times when
I did not have enough grants from other sources. What is more
important is that there are no strings attached to his support and
it is up to the last author to explore what he wants to. CJZ would
also like to thank R. Iengo for encouragements and his interests
in this work, helpful discussions and comments; to C.~F.~Berger,
Z.~Bern, Z.~Chang, L.~J.~Dixon, B.~Feng, D.~Forde, E.~Gava, H. -Y.
Guo, D.~A.~Kosower, K.~S.~Narain, K.~Wu, Y.~-S.~Wu, Z.~Xu and
Z.~-X.~Zhang for discussions and comments; to Prof. X. -Q. Li and
the hospitality at Nankai University where we can have good food;
and finally to Prof. S. Randjbar-Daemi and the hospitality at
Abdus Salam International Center for Theoretical Physics, Trieste,
Italy. This work is supported in part by funds from the National
Natural Science Foundation of China with grant number 10475104 and
10525522.

\end{document}